\documentclass[12pt]{article}
\input epsf.sty
\newcommand{\be}{\begin{equation}}
\newcommand{\ee}{\end{equation}}
\newcommand{\ben}{\begin{eqnarray}\displaystyle}
\newcommand{\een}{\end{eqnarray}}

\usepackage{amssymb}
\usepackage{amsmath}
\usepackage{comment}
\usepackage{mathrsfs}
\usepackage[all]{xy}
\usepackage[dvips]{graphicx}
\usepackage{epsfig,color}
\def\be{\begin{equation}}
\def\ee{\end{equation}}
\def\ba{\begin{align}}
\def\ea{\end{align}}

\makeatletter
\@addtoreset{equation}{section}

\makeatletter
\renewcommand\section{\@startsection {section}{1}{\z@}%
                                   {-3.5ex \@plus -1ex \@minus -.2ex}
                                   {2.3ex \@plus.2ex}%
                                   {\normalfont\large\bfseries}}
\renewcommand\subsection{\@startsection{subsection}{2}{\z@}%
                                     {-3.25ex\@plus -1ex \@minus -.2ex}%
                                     {1.5ex \@plus .2ex}%
                                     {\normalfont\bfseries}}

\begin{document}

\baselineskip18pt

\begin{center}
{
\large{\bf  Is the horizon of an eternal black hole really smooth? }}

\vspace{6mm}

\textit{Nissan Itzhaki}
\break \break
 School of Physics and Astronomy, Tel Aviv University\\ Ramat Aviv 69978, Israel\\ {\it and}\\
 School of Natural Sciences, Institute for Advanced Study\\ 1 Einstein Drive, Princeton, NJ 08540 USA\\
 \textit{E-mail:} \texttt{nitzhaki@post.tau.ac.il}

\vspace{5mm}

\begin{abstract}

\noindent

We point out that in many eternal black holes, including a Schwarzschild eternal  black hole  and an eternal black hole  in $AdS_5$, instant folded strings   are created  in the past wedge and render  the region just outside the horizon singular. We also make a conjecture regarding instant folded D-branes and discuss their possible implications for eternal black holes.
In particular, we argue  that  the bulk modes responsible for Poincare recurrence, when it occurs in the dual quantum field theory,
are either instant folded strings or instant folded D-branes.

\end{abstract}

\vspace{0mm}

\end{center}

\baselineskip=18pt



\section{Introduction}

The fact that the horizon of a large  Eternal Black Hole (EBH) appears to be  regular   at the quantum level \cite{Israel:1976ur} plays a key role also in modern debates about the BH information puzzle. 
In particular,  in the context of  the AdS/CFT correspondence, it was pointed out in \cite{Maldacena:2001kr} that the late time behavior of the two-point function provides a neat realization of the BH information puzzle. On the one hand, the smoothness of the horizon associated with an EBH in $AdS_5$ suggests that the   two point function decays forever. On the other hand, the discrete spectrum of the CFT (when considering on $S^3\times R_t$) implies that the  two point function cannot decay forever and that  Poincare recurrence takes place. 

It was proposed in \cite{Maldacena:2001kr} that a subleading saddle on the gravity side could resolve this puzzle. This proposal was challenged in \cite{Barbon:2003aq} where it was shown that   subleading saddles are not sufficient     to explain the expected time dependence of the Poincare recurrences - a stringy structure just outside the horizon of the EBH in $AdS_5$ is required.  So far no evidence of such a structure has been found.  Here we attempt  to fill this gap and describe the elusive stringy structure foreseen by the authors of \cite{Barbon:2003aq}.

\section{ The basic idea}\label{sone}

The  near horizon region of a large EBH is well described by
\be\label{mi}
ds^2=-du dv,~~
\ee
where, for the time being,  we ignore the angular directions and $v=t+x,~u=t-x$.   This is  a two dimensional  Minkowski space in which  nothing special happens at the horizons, $u=0$ and $v=0$.  
Since, for a large EBH, corrections to (\ref{mi}) are small it is natural to susspect that they cannot affect much the  near  horizon physics. 
Nonetheless, we wish to argue, that there are corrections to  (\ref{mi}), that  at first glance seem to be harmless, but in fact render the EBH horizon singular.

The simplest correction of this type is when the 
 dilaton, $\Phi$, is not constant and  near the horizon it takes the form  
\be\label{dil}
\Phi=\Phi_0-\epsilon uv,~~~
\ee
with a positive $\epsilon$.  Such a dilaton profile is quite common in string theory. For example,
 due to $\alpha^{\prime}$ corrections this is the case  also  in  eternal Schwarzschild BH   \cite{Callan:1988hs, Chen:2021qrz}, and   in EBH in $AdS_5$ \cite{Gubser:1998nz}\footnote{I thank J. Maldacena for reminding me of this paper.} (see also  \cite{Pawelczyk:1998pb}).  
 
Despite the fact that for a large EBH $\epsilon \ll 1$ we attempt to claim now that no matter how small $\epsilon$ is,  as long as it is positive, it   affects dramatically the horizon. We begin with the following  observation.
In the past wedge, $u,v<0$, the dilaton gradient, $\partial_{\mu}\Phi$, is time-like and  points towards the future, and in the future wedge it points towards the past. Similarly, in the right wedge $\partial_{\mu}\Phi$ is space-like and points to the right, and in the left wedge it points to the left (see Fig. 1).    The dilaton gradient is small when $\epsilon\ll1 $ and  it is hard to imagine that this trivial observation can render the horizon singular. 

The point is that a small dilaton gradient can trigger effects that simply do not exist  in its absence. 
Since this is the key point we discuss it in detail.
We start with the simplest case of  a space-like linear dilaton direction with an extra time direction
\be\label{bac}
ds^2=-dt^2+dx^2,~~~~\Phi=Qx,
\ee
for which there is an exact CFT description. 
The dilaton gradient, $Q$, does not affect the equations of motion, but it does modify the Virasoro constraints in an interesting way - it adds a linear term
\be\label{ld}
-(\partial_{\pm}t)^2+ (\partial_{\pm}x)^2-Q\partial_{\pm}^2 x=0.
\ee
The dilaton gradient term, $Q\partial_{\pm}^2 x$, is subleading in the  $\alpha^{'}$ expansion (we work with $\alpha^{'}=1$) compared 
to the standard term $ (\partial_{\pm}x)^2$, but since  it is  linear in $x$ it can dominate the constraints and introduce novel features that are simply absent when $Q=0$.

In the case of (\ref{bac}) the new feature is a long folded string  \cite{Maldacena:2005hi}
\be\label{lfs}
t=t_0+\tau, ~~~~x=x_0-Q\log\left(\frac12(\cosh(\tau/Q)+\cosh(\sigma/Q))\right),
\ee
that does not exist when $Q=0$.\footnote{The 2D `yo yo' solution of \cite{Bardeen:1975gx,Bardeen:1976yt} does not satisfy the Virasoro constraints at the fold  (see \cite{Attali:2018goq} for recent discussion). As a result its 2D YM realizations involve new degrees of freedom at the fold. A recent example,  in the zig-zag model \cite{Donahue:2019fgn}, are the adjoint quarks. For earlier discussion see {\it e.g.} \cite{Ganor:1994rm}.}
 $\tau$ and $\sigma$ are the world sheet coordinates with a range  $-\infty <\tau, \sigma <\infty$. 
The solution describes a string that is stretched from weak coupling, $x=-\infty$, to a finite value of $x$ where it folds (at $\sigma=0$)
\be
x_{fold}(t)=x_0-Q\log\left(\frac12(1+\cosh\left( \frac{t-t_0}{Q}\right) \right),
\ee
and stretched back to weak coupling. $x_{fold}(t)$ is following a time-like trajectory, that is approaching  null trajectories at $t\to\pm \infty$. The null trajectory is right (left) moving for $t< (>) 0$.  The time it takes $x_{fold}(t)$ to turn is of the order of $Q$.  It is during this time that the dilaton gradient term  dominates the standard term  in the Virasoro constraints. 

\vspace{-0mm}
\begin{figure}[h]
	\centering
\includegraphics[scale=0.4]{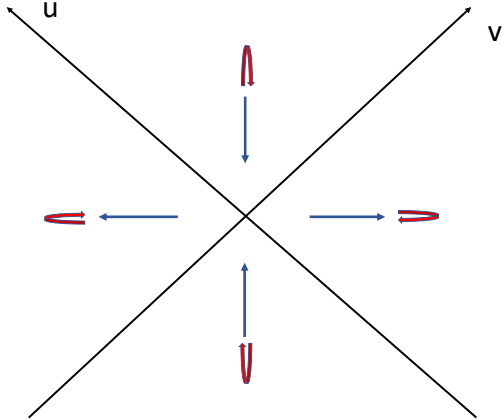}
	\vspace{-10mm}\caption{The string fold topology for $\epsilon>0$. The blue arrows represent the dilaton gradient and the red curved arrow the possible string fold direction.   The string folds point in all wedges  towards the horizon. }
\end{figure}

For $0<Q\ll 1$, that is relevant when $\epsilon\ll 1$, the target-space energy-momentum tensor associated with the long folded string solution takes a particularly simple form that reveals its properties:
\be\label{zx}
T_{uv}=\frac{1}{2\pi}\Theta(v-v_0)\Theta(u-u_0),
\ee
 is due to  the tension  in the bulk of the folded string. Where $\Theta$ is the step function and $v_0=t_0+x_0,~ u_0=t_0-x_0$. At the fold there is  a null flux
\be\label{xz}
T_{uu}=\frac{v_0-v}{2\pi}\Theta(v_0-v)\delta(u-u_0),~~~~T_{vv}=\frac{u-u_0}{2\pi}\Theta(u-u_0)\delta(v_0-v),
\ee
which implies that for $t<t_0$ the null momentum at the fold $P^{v}=(v_0-v)/2\pi$ is positive and decreases with time due to the expansion of the folded string until it vanishes  at the turning point, $t=t_0$. For $t>t_0$ the null momentum at the fold $P^{u}=(u-u_0)/2\pi$  is positive and increases with time as the string shrinks.

The background (\ref{bac}) is a good approximation to (\ref{dil}), when expanding around any point in the left and right wedges, with $Q>(<)0$ in the right (left) wedge. Thus in the left (right) wedge 
strings can fold to the right (left).  

Consider a folded string in the right wedge  that turns at $t_0=0$ and $x_0>0$ (see figure 2(a)). The time scale associated with the turning of $x_{fold}(t)$ is short - it scales like  $ Q=\epsilon x_0$.  In particular, it is much shorter than the scale set by the second derivative of the dilaton and the curvature, $1/\epsilon$.  This means that  for $t< -Q$ and for $t>Q$  
a good approximation to $x_{fold}(t)$ is   a null trajectory.  For $t<-Q$ the null trajectory is  right moving and for $t>Q$ it is left moving. 
Therefore, without knowing the exact folded string solutions in this background, we can tell  that in the right wedge there is a folded string solution that is well approximated by  figure 2(a). In the left wedge there is a mirror solution.  

Even without knowing the extension of the folded string in figure 2(a)  to the other wedges it is clear
that its energy   is at least of the order of $x_0/\pi$. To trust the classical solution we need $x_0\gg 1$ which means that $E\gg 1$ and that these strings are irrelevant at the IR. In particular, they cannot render the horizon singular.

\vspace{-0mm}
\begin{figure}[h]
	\centering
\includegraphics[scale=0.4]{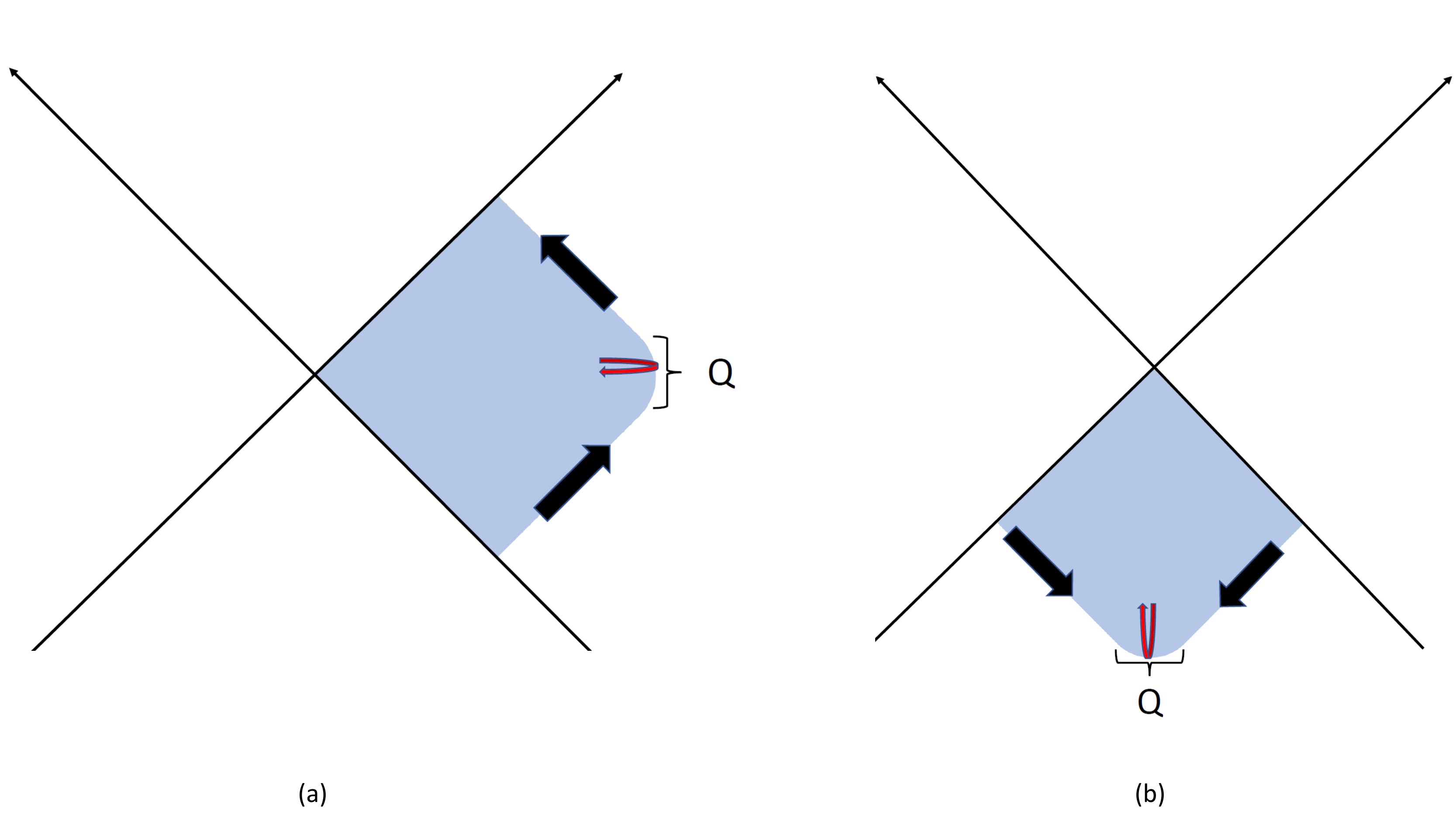}
\vspace{-1mm}	\caption{A folded string in the right wedge (a) and in the past wedge  (b). Both are consistent with the fold topology depicted in figure 1 and in both the folds  asymptote  to  null trajectories in a  short time  that scales like $Q$. The null momenta at the folds, marked with  black arrows, is pointing to the future (past) in the space (time) like case.   }
\end{figure}

The dilaton gradient triggers more extreme  effects  when it is time-like and points to the future. Again, it is instructive to consider the constant dilaton gradient case first
\be\label{bact}
ds^2=-dt^2+dx^2,~~~\Phi=Qt, 
\ee
with $Q>0$. Now the Virasoro constraint are
\be\label{ldb}
-(\partial_{\pm}t)^2+ (\partial_{\pm}x)^2-Q\partial_{\pm}^2 t=0.
\ee
and the dilaton slope term, $Q\partial_{\pm}^2 t$, triggers the creation of an Instant Folded String (IFS) that are described by \cite{Itzhaki:2018glf}
\be\label{ifs}
x=x_0+\sigma,~~~t=t_0+Q\log\left(\frac12(\cosh(\tau/Q)+\cosh(\sigma/Q))\right).
\ee 

While technically it appears similar to the folded string solution (\ref{lfs}) the physical process it describes
is quite different.  
What (\ref{ifs}) describes is a closed folded string that is created {\it classically} at size zero, at $x=x_0$ and $t=t_0$, and  is expanding rapidly. The  fold, located at $\tau=0$, is following a space-like trajectory
\be
t=t_0+Q\log\left(\frac12(1+\cosh\left( \frac{x-x_0}{Q}\right) \right),
\ee
 that very quickly, at time scales of the order of $Q$, asymptotes to a null trajectory.

IFSs are more extreme than the long folded strings since their energy vanish,  hence they can modify the IR physics dramatically.  The energy of an IFS must vanish since it did not exist before $t_0$ and since, from a fundamental string point of view, the background (\ref{bact}) is  time translation invariant.
 The way  the IFS's energy  vanishes is interesting. As in the space-like case, the energy density in the bulk of the IFS is positive due to the string tension. This positive energy is canceled against negative energy at the fold \cite{Attali:2018goq}. As the IFS expands the energy at the folds decreases in such a way that the total energy remains zero.  This is reflected in the  energy-momentum tensor which, again, 
 in the limit $Q\ll 1$,   takes a particularly simple form
\ben
T_{uv}&=&\frac{1}{2\pi}\Theta(v-v_0)\Theta(u-u_0),\\ \nonumber
T_{uu}&=&\frac{v_0-v}{2\pi}\Theta(v-v_0)\delta(u-u_0),~~~T_{vv}=\frac{u_0-u}{2\pi}\Theta(u-u_0)\delta(v-v_0).~
\een
Just like in the space-like case, $T_{uv}$ describes the positive tension of the folded string. The  difference is that now the null momenta at the folds $P^v=(v_0-v)/2\pi$ and $P^u=(u_0-u)/2\pi$ are negative and  decrease with time due to the expansion of the folded string.

The background (\ref{ldb}) is a good approximation to  the background (\ref{dil})
  when expanding around any point in the past and future wedges, with $Q>0$ in the past wedge and $Q<0$ in the future wedge. 
 This means that in the past (future) wedge 
string can fold to the future (past). 
Consider an IFS that is created in the past wedge at $x_0=0$ and $t_0<0$. The time it takes $x_{fold}$ to approach a null trajectory is short - it scales like  $ Q=-\epsilon t_0$. Hence, just like in the space-like case, without knowing the exact folded string solution we conclude that in the past  wedge there is an IFS that looks like in figure 2(b). In the future wedge there is a time-reversal solution - a  closed folded strings that shrinks and disappear at $x_0=0$ and $t_0>0$.

To leading order in $\epsilon$ the energy associated with an IFS that is created in the past wedge vanishes. The energy is not identically zero since unlike (\ref{bac}), the background (\ref{dil}) is not invariant under time translation. Hence we expect the exact IFS solution to acquire with time a small energy that scales like $\epsilon$. That is, like in (\ref{bac}), the IFS is created at zero size with vanishing energy, only that now  its energy  grows with time  $E\sim \epsilon (t-t_0)$.

So far we discussed the possible folded  string solutions in each wedge separately. Now we  describe solutions that are valid in all wedges. 
A natural guess is the  configuration in figure 3(a) which is the most symmetric configuration  consistent with the fold topology.  
\vspace{-0mm} 
 \begin{figure}[h]
	\centering
\includegraphics[scale=0.4]{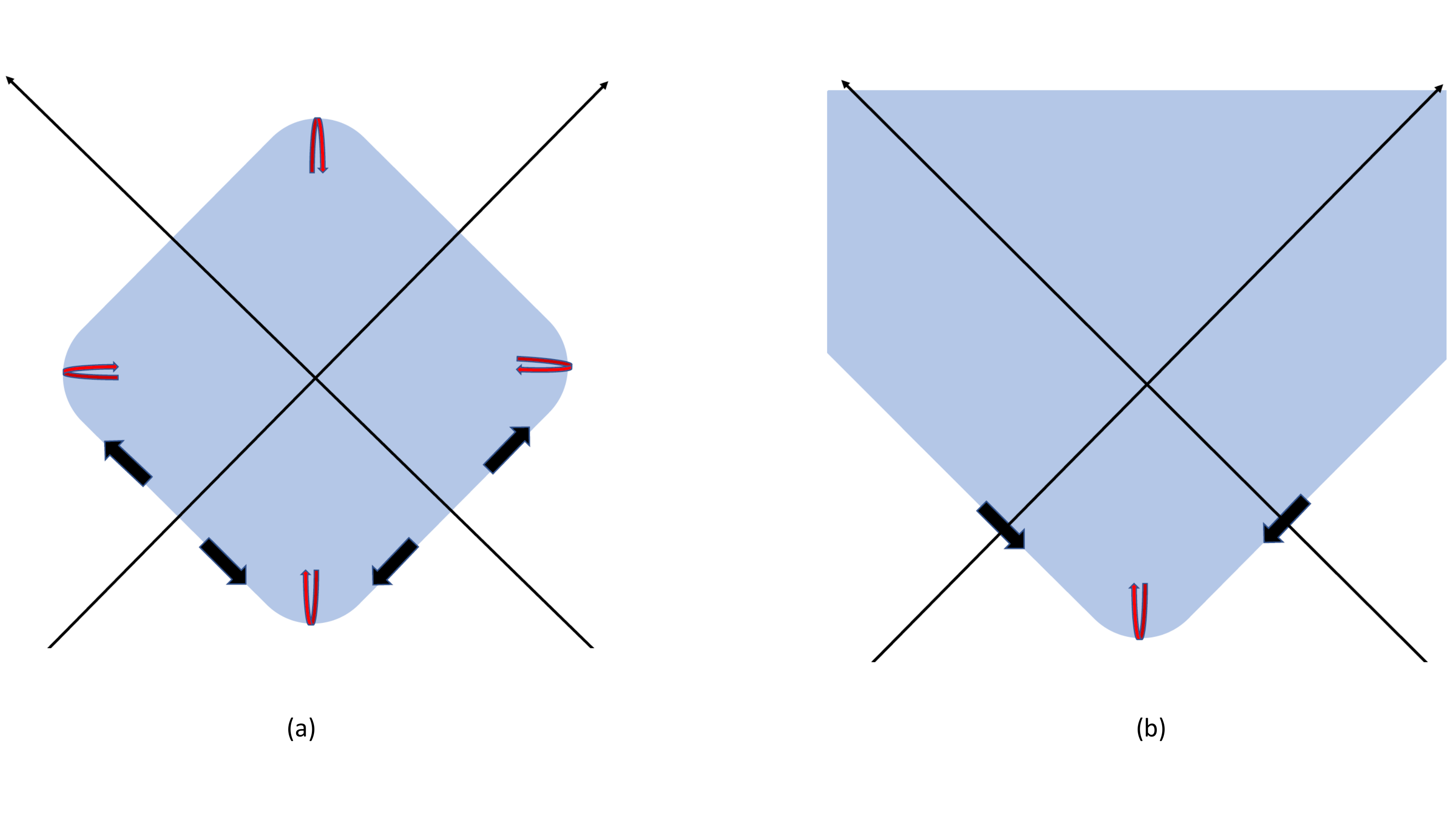}
\vspace{-10mm}	\caption{(a) A natural guess for the folded string configuration  that is consistent with the fold topology. The null momenta at the folds imply that this guess is not consistent with energy conservation.  (b) The consistent configuration in which the null momenta at the folds always point to the past. }
\end{figure} 
There are, however, simple worldsheet and target space arguments that show that this configuration  is inconsistent.  
The worldsheet argument\footnote{I thank E. Witten for this argument.} is that it  has a topology of a sphere  while we work in Lorentzian signature.  Any attempt to reconcile these facts will generate a singularity.  The target-space problem  is that it 
violates  energy-momentum conservation.  To see this recall that in the past wedge the null momenta at the folds are negative (they point to the past). When the folds cross the horizons and enter the left and right wedges the string is still growing which means, by energy conservation, that the null momenta have to become even more negative. In particular, they cannot vanish which, as discussed above, is necessary  for  the folds to  turn sharply  and form the symmetric configuration of figure 3(a). Classically its only option is to continue growing as described in figure 3(b).

An IFS is not going to expand indefinitely when interactions are taken into account.  
To estimate the IFS lifetime  we follow the semi-classical approach of  \cite{Iengo:2006gm}, which for a large spinning folded string  agrees with  the exact CFT calculations \cite{Chialva:2003hg,Iengo:2006gm}. In this approach, away from the fold, the folded string is viewed  as two open strings on top of each other. For the folded string to split the two open strings should split, with a rate like in \cite{Dai:1989cp}, a stringy distance from each other. This 
implies that the lifetime and maximal size of an IFS is of the order of $1/g$.\footnote{In the case of the infinite  long folded string of \cite{Maldacena:2005hi} the string coupling vanishes exponentially fast at infinity.
As a result   this estimate implies a bound on how close to the strong coupling region the folded string can get. This could have implications for \cite{Betzios:2022pji,Ahmadain:2022gfw}.} 

The details of the IFS decay should be interesting to explore since, as discussed below, they  could provide a microscopic description of Hawking radiation. We, however, are in no position to do so since the starting point of such a study is the exact IFS solution which we do not know. Fortunately, for the main point here, it is sufficient to   consider the IFS in its minimal form - an approximate  triangle of size of order $1/g$ (see figure 4(a)).  A more detailed description of  IFSs, that goes beyond their minimal form,  can only  increase their effect.  Hence the discussion below is a lower bound on the effect IFSs can have on the EBH.

The maximal distance from the horizon an IFS, in its minimal form,  can reach is of the order of $1/g$ (see figure 4(a)).
 Still since the production rate of IFSs scales like $Q^2$ \cite{Hashimoto:2022dro}, naively their effect  is negligible for a large EBH (with $\epsilon\ll 1$). However, the EBH is eternal and even a tiny production rate can, in principle, generate an infinite effect. This is the case with  IFSs since they expand basically at the speed of light. 
 Concretely, since  the EBH  is boost invariant  there are infinitely  many  IFS configurations that are related to the one in figure 4(a) by a boost. Hence for any positive $\epsilon$, an  observer that attempts to cross the EBH horizon will encounter an infinite number of IFSs (see figure 4(b)).  This is the sense in which the EBH horizon is singular.   
\vspace{-0mm}
\begin{figure}[h]
	\centering
\includegraphics[scale=0.5]{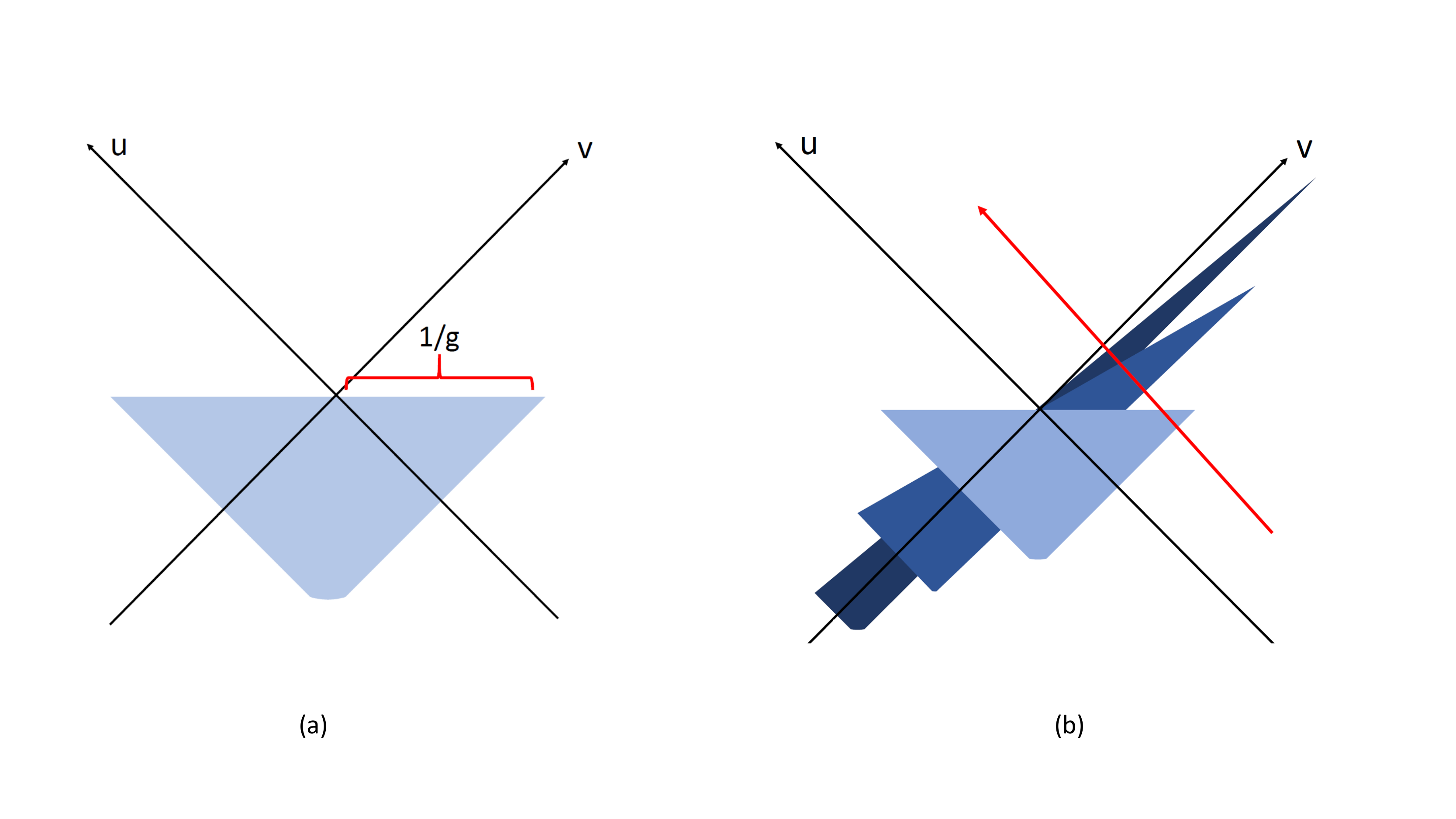}
\vspace{-10mm}	\caption{When interaction are taken into account the size and lifetime of an IFS is of order $1/g$. (a) The IFS (in its minimal form) that penetrates the deepest into the right wedge. (b) Just before crossing the horizon an infalling observer, represented by the red arrow, crosses an infinite amount of IFS that are related to the one in (a) by a boost. 
}
\end{figure}

The conclusion that the horizon is singular is robust and is not sensitive to the initial condition. A natural initial condition, in the spirit of the nice-slice argument \cite{Lowe:1995ac}, is  that there are no IFS at some invariant distance, $\rho\gg 1$, from the past singularity. Such an initial condition is natural since it does not break the boost invariance of the EBH, and this slice is nice since both the curvature and string coupling are small for $\rho\gg 1$.  Since the IFSs that are relevant  to the discussion above are created close to the horizon this initial condition does not affect the mechanism discussed above.   Moreover, since IFSs are created  at the classical level  their production rate scales also  like $1/g_s^2$ \cite{Hashimoto:2022dro}. Therefore,  this mechanism  is  in fact classical and interactions among  IFSs are strong.

In summary the picture that seems to emerge is that everywhere in the past wedge   IFSs are created to form an IFS condensate. IFSs that are created near the horizon  manage to penetrate a bit into  the left and right wedges. The IFS condensate is hot and radiates what to an observer at infinity looks like Hawking radiation. In $AdS$ this radiation naturally  falls back to the IFS condensate.  The analog of the Hartle-Hawking state \cite{Hartle:1976tp} is such that the condensate  inside the future wedge is dominated by strings that look like the time-reversal of IFS -  closed folded strings that shrinks and disappear at an instant. See figure  5.

\vspace{5mm}
\begin{figure}[h]
	\centering
\includegraphics[scale=0.3]{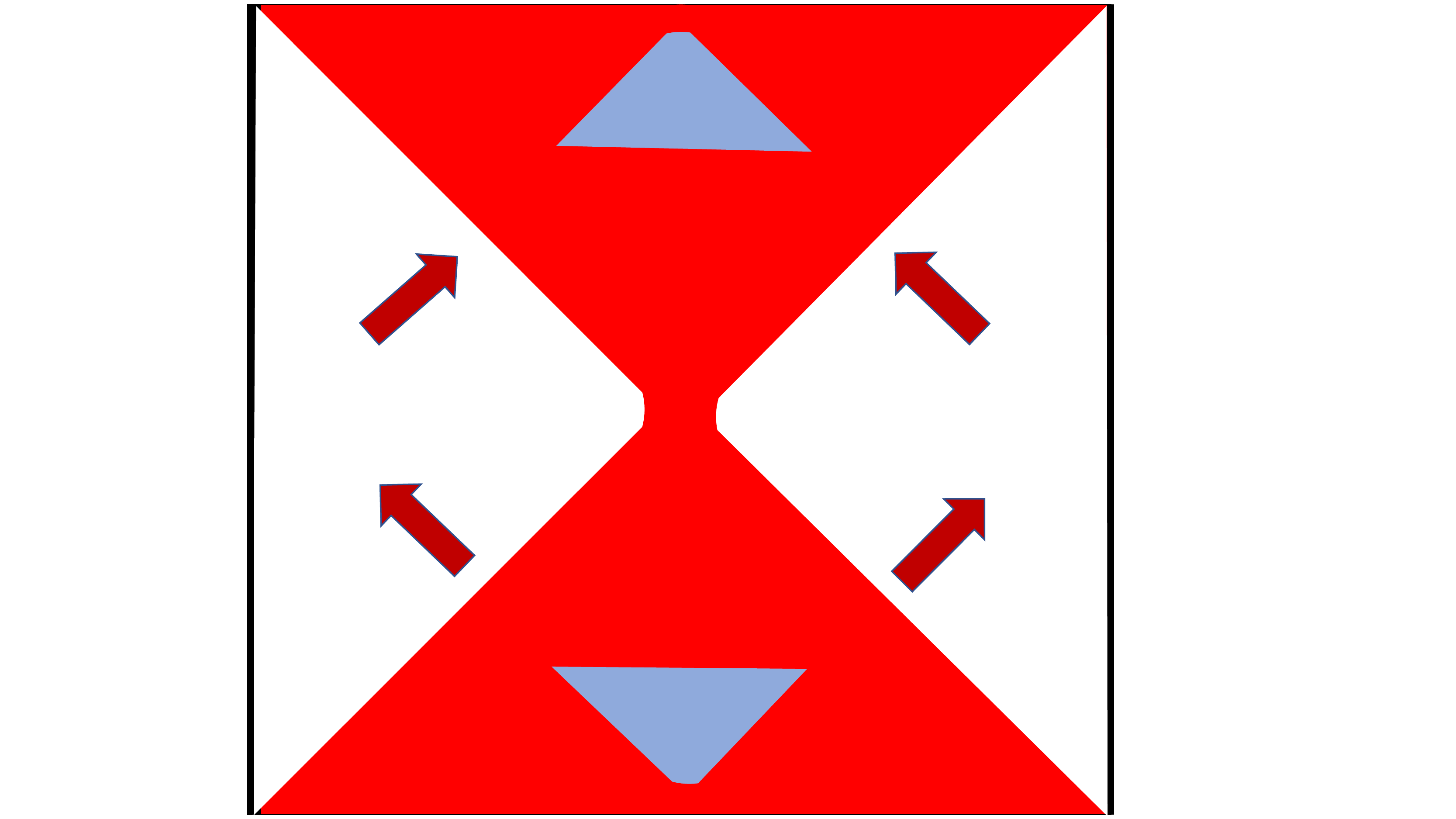}
\vspace{5mm}	\caption{The  proposed Penrose diagram associated with EBH in $AdS_5$.  The past and future wedges as well as the region just outside the horizon are replaced by an IFS condensate. In the past (future) wedge the IFS,  that are represented by 
the triangles, fold towards the future (past).  The condensate emits radiation that bounces back from the boundary. 
}
\end{figure}

\section{Some issues}\label{stwo}

There are  some issues with this mechanism  that we would like to raise.
An obvious issue is the reliance on IFSs.  IFSs are non-standard stringy excitations
 - they are created classically in an instant, and they violate the averaged null energy condition - 
and therefore arguably not part of string theory.  

For several reasons we think that, as strange as they might appear,  IFSs are an integral part of string theory.  First,  as a classical solution to the worldsheet equation of motion and Virasoro constraints they are as good as any other classical solution. Second, in the simplest setup of time-like linear dilaton they  have an exact  worldsheet description \cite{Hashimoto:2022dro} which can be used to calculate their interactions with standard stringy modes and among themselves. A quantity that was actually   calculated in   \cite{Hashimoto:2022dro}, and is used here, is  their production rate. Third, due to the large amount of symmetry associated with the  $SL(2)/U(1)$ EBH, the $(1, 1)$ operator associated with the exponentially small tail of an IFS far from  the horizon was identified in \cite{Giveon:2019gfk,Giveon:2019twx} - its profile exactly matches   semi-classical expectations.

Other, more subtle,  issues:\\
{\bf 1.}  The  mechanism is based on the fact that a dilaton gradient induces a term in the Virasoro constraints 
\be\label{fg}
\partial_{\mu} \Phi \partial_{\pm}^2 x^{\mu},
\ee
that is linear in $x^{\mu}$. Therefore, despite being subleading in the $\alpha^{\prime}$ expansion it  can dominate - even if only for a short while  -  the leading terms in the Virasoro constraints, $g_{\mu\nu} \partial_{\pm} x^{\mu}\partial_{\pm} x^{\nu}$.

At higher orders in $\alpha^{\prime}$  other linear terms  in the Virasoro constraint can appear, and they  can compete with (\ref{fg}). The leading curvature terms  that are linear in $x^{\mu}$, 
$
\partial_{\mu} R \partial_{\pm}^2 x^{\mu}$ and $\nabla_{\nu}R^{\nu}_{\mu} \partial_{\pm}^2 x^{\mu},
$
 vanish in the Schwarzschild EBH background. The leading term that does not vanish is
\be\label{ha}
\nabla_{\mu}R^{\alpha\beta\gamma\delta}  R_{\alpha\beta\gamma\delta}  \partial_{\pm}^2 x^{\mu}. 
\ee
In type II a dilaton gradient is generated only at order  $(\alpha^{\prime})^{3}$ \cite{Gross:1986iv,Chen:2021qrz}. Hence (\ref{ha}) can dominate (\ref{fg}).  This is related to the following issue.

{\bf 2.}  In the case of EBH in $AdS_5$ it is somewhat surprising that what renders the horizon singular are fundamental strings,  that are created only due to a term generated by  $\alpha^{\prime}$ corrections.  Since $AdS_5$ is made  of D3-branes  it is more natural for D3-branes  to play the crucial  role. In section \ref{four}  we argue, subject to a conjecture made in the next section, that, in fact,  this is the case.

{\bf 3.} It also seems strange that the effect depends so dramatically on the sign of $\epsilon$. Clearly, there are EBHs in string theory with $\epsilon<0$.    The effect of  IFSs on such EBHs is much less dramatic since they are created only in the future wedge, and  an infalling observer will encounter only a finite number of them and only after crossing the horizon.\footnote{At least in the case of near extremal NS5-branes this number seems to be related to the Bekenstein-Hawking entropy \cite{Hashimoto:2022dro}.}   Moreover, IFSs that are created classically in the future wedge cannot induce  Poincare recurrence since they cannot render the spectrum of fluctuations outside the BH discrete.

In the next section we consider a setup  designed to address  point 1 above. This setup also suggests a conjecture which, if correct,  addresses points 2 and  3.

\section{A variant of \cite{Maldacena:2001kr}}

 In this section we consider a variant of \cite{Maldacena:2001kr}: the thermofield double state associated with two large $N$ two dimensional  SYM theories with 16 super-charges.  To have a discrete spectrum and Poincare recurrence we  compactify the special direction, $x\sim x+2\pi R$,  with a large $R$ (compared to the scales discussed below).  
We start with a  review of the  conjectured phases of this theory \cite{Itzhaki:1998dd}, before discussing aspects of the thermofield double state.

\subsection{A  review of large $N$  SYM in 2D}

The theory is super-renormalizable as is reflected by the fact that the `t Hooft coupling, $\lambda=g_{YM}^2 N$, has dimension two.   Consequently, the theory is free at the UV, and the effective, dimensionless coupling constant
$
\lambda_{eff}=\frac{\lambda}{E^2},
$
is of order 1 at energies of the order of $\sqrt{\lambda}$. 

For energies much smaller than $\sqrt{\lambda}$ the system is best described by string theory in the near horizon geometry of $N$ D1-branes 
\ben\label{done}
ds^2&=& \frac{U^3}{\sqrt{\lambda}}(-dt^2+dx^2)+\frac{\sqrt{\lambda}}{U^3}dU^2 +\frac{\sqrt{\lambda}}{U}d\Omega_6^2, \\ \nonumber
e^{\Phi}&=&\frac1N \frac{\lambda^{3/2}}{U^3},
\een
where
 as in \cite{Maldacena:1997re}, $U=r/\alpha^{\prime},$ is the energy scale associated with the radial direction $r$, and 
we neglected factors of order $1$.  As usual \cite{tHooft:1973alw} the string coupling scales like $1/N$ and is small in the large $N$ limit. The curvature (in string units)  scales like $R\sim \frac{U}{\sqrt{\lambda}}\sim \frac{1}{\lambda_{eff}^{1/4}}$ and so  when the perturbative description breaks down the SUGRA description takes over. 

As we go further to the IR the string coupling constant becomes large and for $U< \sqrt{\lambda}/ N^{1/3}$ the system is described via the  S-dual background associated with the near horizon limit of $N$ fundamental string
\ben\label{fonef}
ds^2&=& N\left( \frac{U^6}{\lambda^2}(-dt^2+dx^2)+\frac{1}{\lambda}dU^2 +\frac{U^2}{\lambda}d\Omega_6^2\right), \\ \nonumber
e^{-\Phi}&=&\frac1N \frac{\lambda^{3/2}}{U^3}.
\een

The curvature associated with this background scales like $\frac{\lambda}{N U^2}$ and so eventually in the deep IR, $U< \sqrt{\lambda}/ N^{1/2}=g_{YM}$  this description breaks down and the system is best described  via the orbifold $ (R^8)^N/S_N$  conformal field theory \cite{Dijkgraaf:1997vv,Motl:1997th}, also known as matrix strings theory,  associated with the  motion of the $N$ fundamental strings in the transverse space. The various effective descriptions are summarized  in the figure 6.
\vspace{-0mm}
\begin{figure}[h]
	\centering
\includegraphics[scale=0.3]{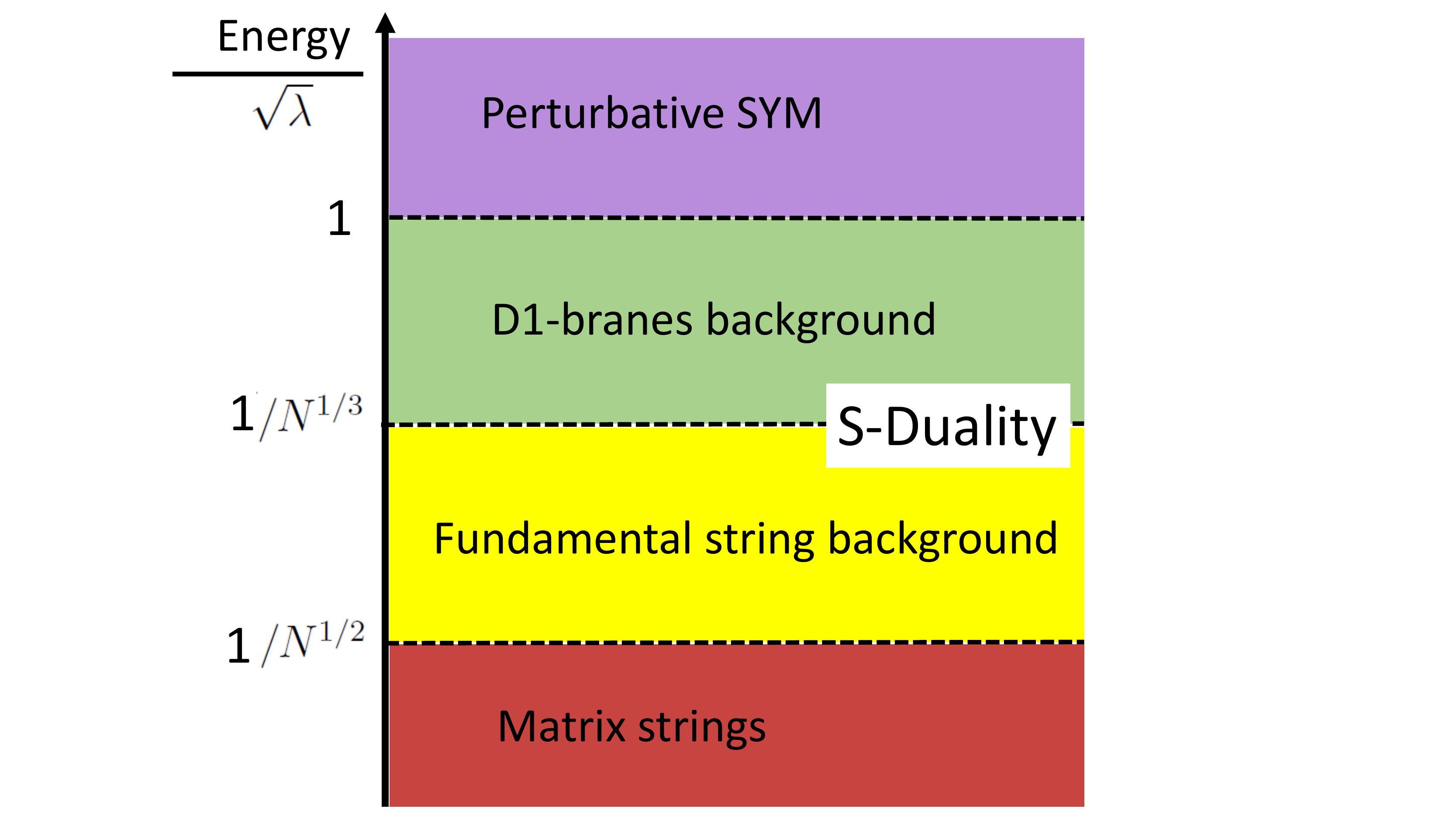}
\vspace{-0mm}	\caption{The conjectured phases of large N SYM in 2D. }
\end{figure}

\subsection{The thermofield double state}

Consider the  thermofield double state with an inverse temperature, $\beta$, that entangles two 2D SYM theories
\be
| TFD \rangle(\beta)=\frac{1}{\sqrt{Z(\beta)}}\sum_{n} e^{-\beta E_n /2}|n\rangle_L\times |n\rangle_R,
\ee
where as usual $|n\rangle_{L,R}$ are the energy eigenstates of the individual theories.

The discussion above implies that the best description of $| TFD \rangle(\beta)$ depends on $\beta$. 
In the  UV, $\beta\ll 1/\sqrt{\lambda}$, the natural description is in terms of the perturbative  degrees of freedom of SYM.
In  the  IR, $\beta\gg \sqrt{N/\lambda}$,  the natural description is in terms  of the perturbative degrees of freedom of the matrix strings theory \cite{Dijkgraaf:1997vv,Motl:1997th} -  the  diagonal elements of the 8 $SU(N)$ matrices, that correspond to the motion of the $N$ strings, and their super partners. 
Since $x$ is compactified in both cases  the spectrum is discrete,  the two point function does not decay forever,  and Poincare recurrence takes place.  

The question is  what happens in the intermediate region
\be
1 \gg \beta \sqrt{\lambda} \gg \sqrt{N},
\ee
where, at least to leading approximation,  the thermofield double state is described in terms of an EBH \cite{Balasubramanian:1998de,Maldacena:2001kr}? 

 In the range
\be\label{fone}
N^{1/3} \gg \beta \sqrt{\lambda} \gg \sqrt{N},
\ee
the relevant EBH is the one associated with $N$ near extremal fundamental strings (the relevant Penrose diagram is depicted in figure 7(a)) and in the range 
\be\label{doner}
1 \gg \beta \sqrt{\lambda} \gg N^{1/3},
\ee
it is  the one associated with $N$ near extremal  D1-branes (the relevant Penrose diagram is depicted in figure 8(a)).  
At the SUGRA level both lead to the standard problem that since the horizon is smooth the spectrum of excitations in its vicinity is continuous \cite{tHooft:1984kcu},  which  implies that   the two point function  decays forever. 

In string theory  the situation is more interesting.   The dilaton in the near horizon  region of the EBH associated with  $N$ near extremal fundamental strings is described by (\ref{dil}) with a small and positive
 $\epsilon$. The discussion  in   section \ref{sone} suggests that the horizon in this case is not smooth but filled with IFSs, 
 which  implies that the spectrum of fluctuations in its vicinity is discrete.  This fits neatly  with    \cite{Barbon:2003aq}  that argue that a  stringy structure just outside the horizon is needed to explain the expected time dependence of the Poincare recurrences. It appears that, at least in the case of the EBH associated with near extremal fundamental strings, the stringy structure anticipated in \cite{Barbon:2003aq} is the IFS condensate. 

Note that, unlike EBH in $AdS_5$, now $\epsilon$ is non vanishing at the SUGRA level. Therefore, the IFS production trigger, $\partial_{\mu}\Phi$, is the leading linear term in the Virasoro constraints  and other possible linear terms, discussed in section \ref{stwo}, that can appear at higher orders in $\alpha^{\prime}$, are negligible. This also addresses the second issue in section \ref{stwo}: now the background is made of fundamental strings and  it is natural that IFSs are the ones that render the horizon singular.

What happens in the D1-branes range (\ref{doner})? In this case $\epsilon <0$ and IFS are  created only in the future  wedge. Such IFSs can modify the BH interior considerably, but they cannot turn  the  spectrum of fluctuations just outside  the horizon discrete. 

A concrete example with $\epsilon <0$ that   
illustrates this  is the EBH associated with  $k$ near extremal NS5-branes \cite{Maldacena:1997cg}. This background has a  coset CFT description \cite{Witten:1991yr,Elitzur:1990ubs,Mandal:1991tz,Dijkgraaf:1991ba} which was used to calculate the exact reflection coefficient, including all perturbative and non-perturbative $\alpha^{\prime}$ corrections, on the sphere  \cite{Teschner:1999ug} (see also \cite{Giribet:2000fy}). Since the production rate of IFSs scales like $1/g^2$ \cite{Hashimoto:2022dro} they are expected to leave their mark on this calculation. In fact, the screening operator used in \cite{Giribet:2000fy} to perform this calculation is the operator that describes the IFS in this background  \cite{Giveon:2019gfk,Giveon:2019twx}. The exact reflection coefficient differs from the SUGRA result only by a phase which implies that IFSs affect very little the region outside the horizon. In particular, the spectrum of fluctuations outside the horizon remains continuous, as is clear from the fact the reflection coefficient  decays exponentially fast for energies larger than $1/\sqrt{k}$. The dependence of this phase on the energy  is highly non-trivial \cite{Giveon:2016dxe}, and it was argued  in \cite{Ben-Israel:2017zyi,Itzhaki:2018rld} that this implies that the region beyond the horizon is not smooth.  This fits well with the fact that IFSs are classically created only behind the horizon \cite{Itzhaki:2018glf}.

\vspace{-0mm}
\begin{figure}[h]
	\centering
\includegraphics[scale=0.4]{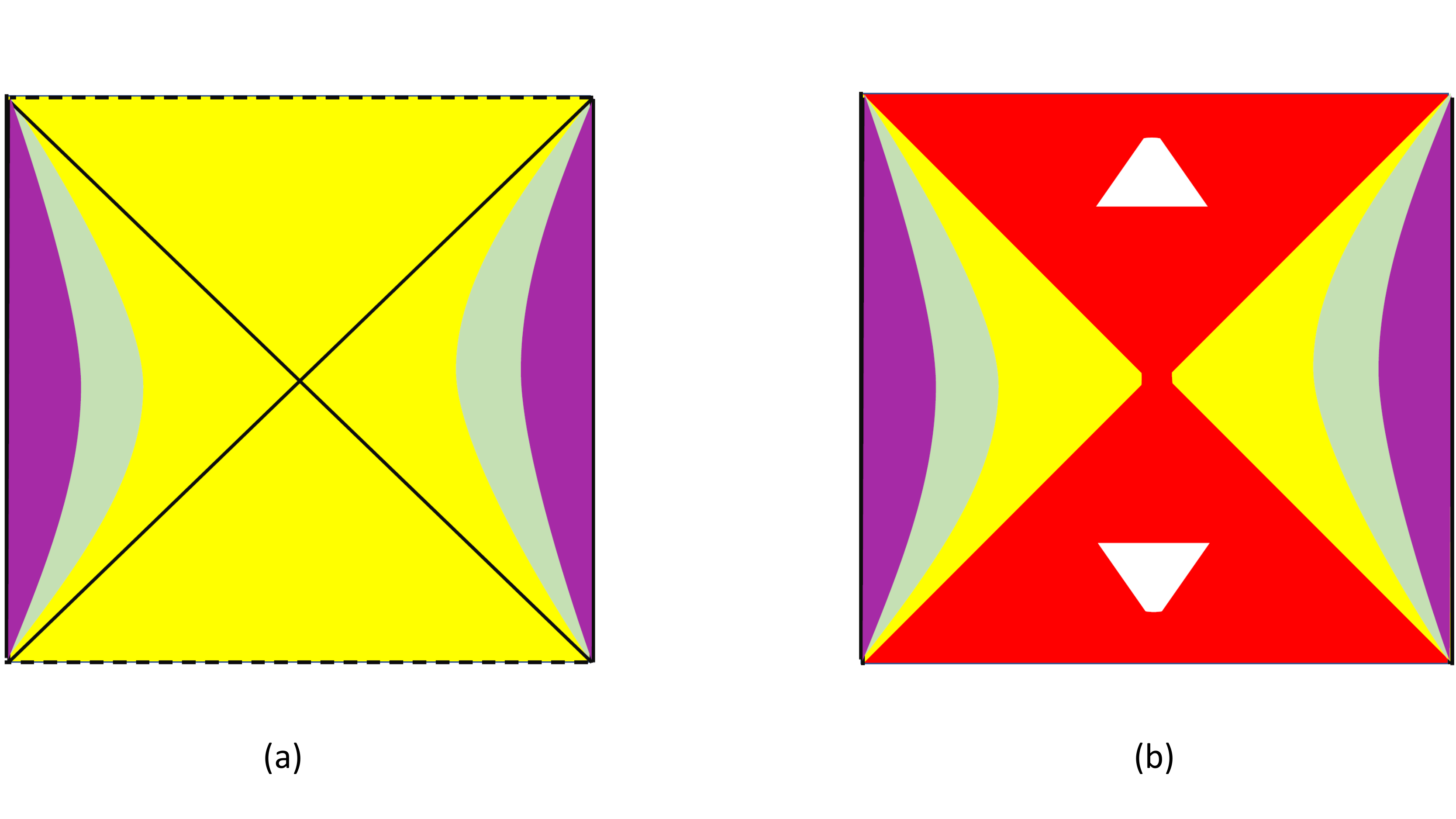}
\vspace{-0mm}	\caption{(a) The standard Penrose diagram associated with the eternal near extremal fundamental strings. The purple regions have a large curvature and are best described by perturbative SYM. The dashed lines represent the EBH singularities. The green regions are described by the near extremal D1-branes background and the yellow by the near extremal F1 background. 
Both the curvature and string coupling are small at the horizons.  (b) Since $\epsilon$ is positive the region just outside the horizon as well as the past and future wedges are replaced by an IFS condensate (marked in red). 
The white triangles indicate the dominant shape of the IFSs that form the IFS condensate.    }
\end{figure}

Even if we put aside field theory considerations, from a pure bulk perspective it seems highly  peculiar that in a certain temperature range  the spectrum of fluctuations near the horizon is continuous and  in a nearby temperature range it is  discrete. A     way to evade this peculiarity is to argue that since the EBH associated with $N$ near extremal D1-branes
is S-dual to the EBH associated with $N$ near extremal fundamental strings its horizon is not smooth either, this time due to creation in the past wedge of  Instant Folded D1-branes (IFD1-branes) - the naive S-dual of  IFSs.
Since IFSs are not BPS it is not clear how they transform under S-duality and the existence of an IFD1-branes is  a conjecture.\footnote{We hope that this conjecture is provable.  Linear dilaton CFT admits several D-branes that are absent when the dilaton slope vanishes \cite{Fateev:2000ik,Teschner:2000md,Lukyanov:2003nj}. Not all of them are easily described by the DBI action. The space like version of this conjecture is that on top of these in the background (\ref{bac}) there are long folded D1-branes - the S-dual of (\ref{lfs}) with a shape roughly described by (\ref{lfs}), with $Q\to -Q$ and $x\to -x$.    } The  precise conjecture is:
 \begin{quote}
{\it A time-like dilaton gradient that points to the past triggers the creation of IFD1-branes with a shape roughly described by (\ref{ifs}), with $Q\to -Q$ and $t\to -t$.} 
 \end{quote}
The expected lifetime of an IFD1-brane is of the order of the string scale.  The reason is that  away from the fold an IFD1-brane looks like a D1-brane on top of an anti D1-brane - a system  which admits  an open string tachyon with $m^2\sim -1$ \cite{Sen:1998sm}.  We do not know the production rate of IFD1-branes but it is reasonable to suspect that it is finite, in which case the eternity of the eternal near extremal D1-branes background suggests that they should render the near horizon spectrum discrete.

\vspace{-0mm}
\begin{figure}[h]
	\centering
\includegraphics[scale=0.4]{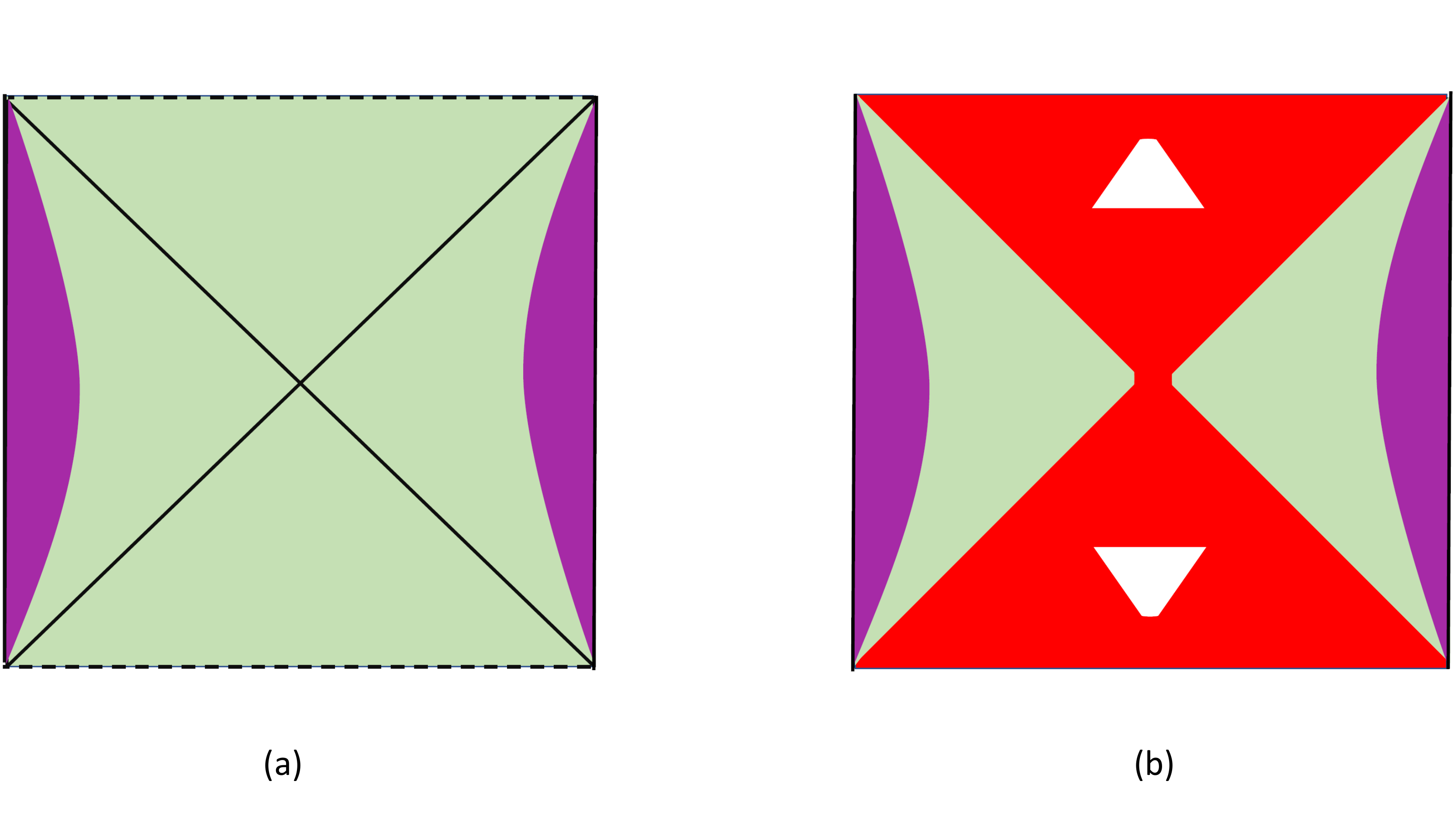}
\vspace{-0mm}	\caption{(a) The standard Penrose diagram associated with  eternal near extremal D1-branes. The purple regions have a large curvature and are best described by perturbative SYM. The dashed lines represent the EBH singularities. The green regions are described by the near extremal D1-branes background. 
Both the curvature and string coupling are small at the horizons. (b) As $\epsilon$ is negative we conjectured that the region just outside the horizon as well as the past and future wedges are replaced by an IFD1-brane condensate (marked in red). 
The white triangles indicate the dominant shape of the IFD1-branes that form the IFD1-brane condensate. 
   }
\end{figure}

\section{Back to Schwarzschild and $AdS_5$}\label{four}

In the previous section we conjectured that a time-like dilaton gradient that points  to the past  
 triggers  the creation of  IFD1-branes. In this section we assume  the conjecture is correct  and apply T-duality to find triggers for the creation of other instant folded D-branes.  We discuss possible implications to the Schwarzschild EBH and the EBH in $AdS_5$.

  Consider a setup in which $\partial_t \Phi(t)<0$ and one of the  directions is compactified with a radius that depends on $t$,    $y\sim y+2\pi R(t)$.  Since the creation of an IFD1-brane, that is extended in $t$ and some other direction $x$,  is a local process 
  it is not expected to be sensitive to the fact that $y$ is compactified.
The IFD1-branes evolution, however, is sensitive to $R(t)$, especially when $R<1$. To see this we recall that for $R<1$ there are tachyons, on top of the one discussed in the previous section, due to open strings that  are stretched between  the  IFD1-brane and its images in the covering space. As we decrease $R$ the number of these tachyonic modes grows, which implies that the smaller $R$ is the shorter the lifetime of the IFD1-brane is.

When $R(t)\ll 1$ it is natural to apply a time-dependent T-duality \cite{Smith:1991up} which takes $R(t)\to\tilde{R}(t)=1/R(t)\gg 1$ 
and  the IFD1-brane to  an IFD2-brane
that wraps $\tilde{y}$.
T-duality  also changes the string coupling 
$
g_s=\tilde{g}_s/\tilde{R} $  \cite{Ginsparg:1986wr,Buscher:1987sk},
which means that the trigger for   the creation of the IFD2-brane is a time-like
\be\label{hqh}
\partial_{\mu}(\tilde{r}-\tilde{\Phi}),
\ee
that points to the future where $\tilde{r}$ is the radion field, $\tilde{r}=\log(\tilde{R})$.

This seems to be relevant for an eternal   Schwarzschild BH in
type IIA string theory.  In the past wedge the $S^2$ is growing with time and so there is a $\tilde{r}$
such that $\partial_{\mu} \tilde{r}$ is time-like and points to the future. To show this  explicitly we 
write the background in the familiar form
 \be\label{sch}
 ds^2=-\left(1-\frac{2M}{r}\right) dt^2+\frac{dr^2}{(1-\frac{2M}{r})}+r^2(d\theta^2+\sin^2\theta d\phi^2),~~~~\Phi=\Phi_0. 
 \ee
 The radion associated with the $\phi$ direction,   $\tilde{r}=\log(r \sin\theta)$, is time-like and points to the future in the past wedge when
 \be\label{oas}
 r<2M \sin^2 \theta.
 \ee
Therefore, it  triggers the creation of  IFD2-branes which  wrap  $\phi$ and have a similar shape to the IFS in figure 4(a) in the $u,v$ plane. 
Note that the classical background (\ref{sch}), with no  $\alpha^{\prime}$ corrections, admits  (\ref{hqh}) that is time-like and points to the future. 

The discussion above  suggests that the larger the $S^1$ is the shorter the lifetime of the IFD2-brane is. This combined  with  (\ref{oas}) implies that the lifetime of the IFD2-branes that dominate the near horizon dynamics  is short. Again, the eternity of the eternal  Schwarzschild BH guarantees that as long as this lifetime is finite the horizon is singular, in the sense discussed in section 2.  

In case there is another $S^1$ we can apply T-duality once more to find that the trigger for the creation of IFD3-branes (which wrap  the two cycles)  is a time-like 
\be\label{pqw}
\partial_{\mu}(\tilde{r}_1+\tilde{r}_2-\tilde{\Phi}),
\ee
that points to the future where $\tilde{r}_i$ are the radion fields, $\tilde{r}_i=\log(\tilde{R}_i),~i=1,2$. 

This appears to be relevant for EBH in $AdS_5$. 
To end up with a discrete spectrum on the field theory side we can  consider the theory on $S^3\times R_t$  or on $T^3\times R_t$. In both cases the dual EBH  involves two cycles that are growing with time in   
the  past wedge, and IFD3-branes that wrap these two cycles will render the horizon singular.  Again the classical background of an EBH in $AdS_5$, with no  $\alpha^{\prime}$ corrections, admits  (\ref{pqw}) that is time-like and points to the future.  This means that the red region in figure 4 is more likely to represent an IFD3-branes condensate than IFS condensate.

More generally, the conjecture implies that IFSs are only the tip of the iceberg and that there are many objects in string theory that are created in an instant in time-dependent situations.
These instant objects could play an important role  also in cosmology, where like IFS \cite{Itzhaki:2021scf}, they  are expected to induce negative pressure at no energy cost  - this time when  scalars other than the dilaton  vary with time.  
They may even play a role in extreme situations in astrophysics.

\vspace{5mm}

\section*{Acknowledgments}

I  thank  A. Hashimoto and E. Witten for helpful discussions. I also thank Y. Zigdon for pointing out a typo in (2.6) and (2.12). 
Work  supported in part by the ISF (grant number 256/22), BSF (grant number 2018068) and by the Adler Family Fund.


\begin{thebibliography}{1}


\bibitem{Israel:1976ur}
W.~Israel,
``Thermo field dynamics of black holes,''
Phys. Lett. A \textbf{57}, 107-110 (1976)
doi:10.1016/0375-9601(76)90178-X






\bibitem{Maldacena:2001kr}
J.~M.~Maldacena,
``Eternal black holes in anti-de Sitter,''
JHEP \textbf{04}, 021 (2003)
doi:10.1088/1126-6708/2003/04/021
[arXiv:hep-th/0106112 [hep-th]].




\bibitem{Barbon:2003aq}
J.~L.~F.~Barbon and E.~Rabinovici,
``Very long time scales and black hole thermal equilibrium,''
JHEP \textbf{11}, 047 (2003)
doi:10.1088/1126-6708/2003/11/047
[arXiv:hep-th/0308063 [hep-th]].








\bibitem{Callan:1988hs}
C.~G.~Callan, Jr., R.~C.~Myers and M.~J.~Perry,
``Black Holes in String Theory,''
Nucl. Phys. B \textbf{311}, 673-698 (1989)
doi:10.1016/0550-3213(89)90172-7

\bibitem{Chen:2021qrz}
Y.~Chen,
``Revisiting $R^4$ higher curvature corrections to black holes,''
[arXiv:2107.01533 [hep-th]].


\bibitem{Gubser:1998nz}
S.~S.~Gubser, I.~R.~Klebanov and A.~A.~Tseytlin,
``Coupling constant dependence in the thermodynamics of N=4 supersymmetric Yang-Mills theory,''
Nucl. Phys. B \textbf{534}, 202-222 (1998)
doi:10.1016/S0550-3213(98)00514-8
[arXiv:hep-th/9805156 [hep-th]].

\bibitem{Pawelczyk:1998pb}
J.~Pawelczyk and S.~Theisen,
``AdS(5) x S**5 black hole metric at O(alpha-prime**3),''
JHEP \textbf{09}, 010 (1998)
doi:10.1088/1126-6708/1998/09/010
[arXiv:hep-th/9808126 [hep-th]].


\bibitem{Maldacena:2005hi}
J.~M.~Maldacena,
``Long strings in two dimensional string theory and non-singlets in the matrix model,''
JHEP \textbf{09}, 078 (2005)
doi:10.1088/1126-6708/2005/09/078
[arXiv:hep-th/0503112 [hep-th]].

\bibitem{Bardeen:1975gx}
W.~A.~Bardeen, I.~Bars, A.~J.~Hanson and R.~D.~Peccei,
``A Study of the Longitudinal Kink Modes of the String,''
Phys. Rev. D \textbf{13}, 2364-2382 (1976)
doi:10.1103/PhysRevD.13.2364

\bibitem{Bardeen:1976yt}
W.~A.~Bardeen, I.~Bars, A.~J.~Hanson and R.~D.~Peccei,
``Quantum Poincare Covariance of the D = 2 String,''
Phys. Rev. D \textbf{14}, 2193 (1976)
doi:10.1103/PhysRevD.14.2193

\bibitem{Attali:2018goq}
K.~Attali and N.~Itzhaki,
``The Averaged Null Energy Condition and the Black Hole Interior in String Theory,''
Nucl. Phys. B \textbf{943}, 114631 (2019)
doi:10.1016/j.nuclphysb.2019.114631
[arXiv:1811.12117 [hep-th]].


\bibitem{Donahue:2019fgn}
J.~C.~Donahue and S.~Dubovsky,
``Classical Integrability of the Zigzag Model,''
Phys. Rev. D \textbf{102}, no.2, 026005 (2020)
doi:10.1103/PhysRevD.102.026005
[arXiv:1912.08885 [hep-th]].

\bibitem{Ganor:1994rm}
O.~Ganor, J.~Sonnenschein and S.~Yankielowicz,
``Folds in 2-D string theories,''
Nucl. Phys. B \textbf{427}, 203-244 (1994)
doi:10.1016/0550-3213(94)90275-5
[arXiv:hep-th/9404149 [hep-th]].



\bibitem{Itzhaki:2018glf}
N.~Itzhaki,
``Stringy instability inside the black hole,''
JHEP \textbf{10}, 145 (2018)
doi:10.1007/JHEP10(2018)145
[arXiv:1808.02259 [hep-th]].


\bibitem{Iengo:2006gm}
R.~Iengo and J.~G.~Russo,
``Handbook on string decay,''
JHEP \textbf{02}, 041 (2006)
doi:10.1088/1126-6708/2006/02/041
[arXiv:hep-th/0601072 [hep-th]].

\bibitem{Chialva:2003hg}
D.~Chialva, R.~Iengo and J.~G.~Russo,
``Decay of long-lived massive closed superstring states: Exact results,''
JHEP \textbf{12}, 014 (2003)
doi:10.1088/1126-6708/2003/12/014
[arXiv:hep-th/0310283 [hep-th]].


\bibitem{Dai:1989cp}
J.~Dai and J.~Polchinski,
``The Decay of Macroscopic Fundamental Strings,''
Phys. Lett. B \textbf{220}, 387-390 (1989)
doi:10.1016/0370-2693(89)90892-7




\bibitem{Betzios:2022pji}
P.~Betzios and O.~Papadoulaki,
``Microstates of a $2d$ Black Hole in string theory,''
[arXiv:2210.11484 [hep-th]].

\bibitem{Ahmadain:2022gfw}
A.~Ahmadain, A.~Frenkel, K.~Ray and R.~M.~Soni,
``Boundary Description of Microstates of the Two-Dimensional Black Hole,''
[arXiv:2210.11493 [hep-th]].

\bibitem{Hashimoto:2022dro}
A.~Hashimoto, N.~Itzhaki and U.~Peleg,
``A Worldsheet Description of Instant Folded Strings,''
[arXiv:2209.04988 [hep-th]].

\bibitem{Lowe:1995ac}
D.~A.~Lowe, J.~Polchinski, L.~Susskind, L.~Thorlacius and J.~Uglum,
``Black hole complementarity versus locality,''
Phys. Rev. D \textbf{52}, 6997-7010 (1995)
doi:10.1103/PhysRevD.52.6997
[arXiv:hep-th/9506138 [hep-th]].


\bibitem{Hartle:1976tp}
J.~B.~Hartle and S.~W.~Hawking,
``Path Integral Derivation of Black Hole Radiance,''
Phys. Rev. D \textbf{13}, 2188-2203 (1976)
doi:10.1103/PhysRevD.13.2188



\bibitem{Giveon:2019gfk}
A.~Giveon and N.~Itzhaki,
``Stringy Black Hole Interiors,''
JHEP \textbf{11}, 014 (2019)
doi:10.1007/JHEP11(2019)014
[arXiv:1908.05000 [hep-th]].

\bibitem{Giveon:2019twx}
A.~Giveon and N.~Itzhaki,
``Stringy Information and Black Holes,''
JHEP \textbf{06}, 117 (2020)
doi:10.1007/JHEP06(2020)117
[arXiv:1912.06538 [hep-th]].


\bibitem{Gross:1986iv}
D.~J.~Gross and E.~Witten,
``Superstring Modifications of Einstein's Equations,''
Nucl. Phys. B \textbf{277}, 1 (1986)
doi:10.1016/0550-3213(86)90429-3




\bibitem{Itzhaki:1998dd}
N.~Itzhaki, J.~M.~Maldacena, J.~Sonnenschein and S.~Yankielowicz,
``Supergravity and the large N limit of theories with sixteen supercharges,''
Phys. Rev. D \textbf{58}, 046004 (1998)
doi:10.1103/PhysRevD.58.046004
[arXiv:hep-th/9802042 [hep-th]].


\bibitem{Maldacena:1997re}
J.~M.~Maldacena,
``The Large N limit of superconformal field theories and supergravity,''
Adv. Theor. Math. Phys. \textbf{2}, 231-252 (1998)
doi:10.1023/A:1026654312961
[arXiv:hep-th/9711200 [hep-th]].



\bibitem{tHooft:1973alw}
G.~'t Hooft,
``A Planar Diagram Theory for Strong Interactions,''
Nucl. Phys. B \textbf{72}, 461 (1974)
doi:10.1016/0550-3213(74)90154-0

\bibitem{Dijkgraaf:1997vv}
R.~Dijkgraaf, E.~P.~Verlinde and H.~L.~Verlinde,
``Matrix string theory,''
Nucl. Phys. B \textbf{500}, 43-61 (1997)
doi:10.1016/S0550-3213(97)00326-X
[arXiv:hep-th/9703030 [hep-th]].

\bibitem{Motl:1997th}
L.~Motl,
``Proposals on nonperturbative superstring interactions,''
[arXiv:hep-th/9701025 [hep-th]].

\bibitem{Balasubramanian:1998de}
V.~Balasubramanian, P.~Kraus, A.~E.~Lawrence and S.~P.~Trivedi,
``Holographic probes of anti-de Sitter space-times,''
Phys. Rev. D \textbf{59}, 104021 (1999)
doi:10.1103/PhysRevD.59.104021
[arXiv:hep-th/9808017 [hep-th]].


\bibitem{tHooft:1984kcu}
G.~'t Hooft,
``On the Quantum Structure of a Black Hole,''
Nucl. Phys. B \textbf{256}, 727-745 (1985)
doi:10.1016/0550-3213(85)90418-3



\bibitem{Maldacena:1997cg}
J.~M.~Maldacena and A.~Strominger,
JHEP \textbf{12}, 008 (1997)
doi:10.1088/1126-6708/1997/12/008
[arXiv:hep-th/9710014 [hep-th]].

\bibitem{Witten:1991yr}
E.~Witten,
``On string theory and black holes,''
Phys. Rev. D \textbf{44}, 314-324 (1991)
doi:10.1103/PhysRevD.44.314

\bibitem{Elitzur:1990ubs}
S.~Elitzur, A.~Forge and E.~Rabinovici,
``Some global aspects of string compactifications,''
Nucl. Phys. B \textbf{359}, 581-610 (1991)
doi:10.1016/0550-3213(91)90073-7

\bibitem{Mandal:1991tz}
G.~Mandal, A.~M.~Sengupta and S.~R.~Wadia,
``Classical solutions of two-dimensional string theory,''
Mod. Phys. Lett. A \textbf{6}, 1685-1692 (1991)
doi:10.1142/S0217732391001822

\bibitem{Dijkgraaf:1991ba}
R.~Dijkgraaf, H.~L.~Verlinde and E.~P.~Verlinde,
``String propagation in a black hole geometry,''
Nucl. Phys. B \textbf{371}, 269-314 (1992)
doi:10.1016/0550-3213(92)90237-6

\bibitem{Teschner:1999ug}
J.~Teschner,
``Operator product expansion and factorization in the H+(3) WZNW model,''
Nucl. Phys. B \textbf{571}, 555-582 (2000)
doi:10.1016/S0550-3213(99)00785-3
[arXiv:hep-th/9906215 [hep-th]].


\bibitem{Giribet:2000fy}
G.~Giribet and C.~A.~Nunez,
``Aspects of the free field description of string theory on AdS(3),''
JHEP \textbf{06}, 033 (2000)
doi:10.1088/1126-6708/2000/06/033
[arXiv:hep-th/0006070 [hep-th]].

\bibitem{Giveon:2016dxe}
A.~Giveon, N.~Itzhaki and D.~Kutasov,
``Stringy Horizons II,''
JHEP \textbf{10}, 157 (2016)
doi:10.1007/JHEP10(2016)157
[arXiv:1603.05822 [hep-th]].


\bibitem{Ben-Israel:2017zyi}
R.~Ben-Israel, A.~Giveon, N.~Itzhaki and L.~Liram,
``On the black hole interior in string theory,''
JHEP \textbf{05}, 094 (2017)
doi:10.1007/JHEP05(2017)094
[arXiv:1702.03583 [hep-th]].

\bibitem{Itzhaki:2018rld}
N.~Itzhaki and L.~Liram,
``A stringy glimpse into the black hole horizon,''
JHEP \textbf{04}, 018 (2018)
doi:10.1007/JHEP04(2018)018
[arXiv:1801.04939 [hep-th]].



\bibitem{Fateev:2000ik}
V.~Fateev, A.~B.~Zamolodchikov and A.~B.~Zamolodchikov,
``Boundary Liouville field theory. 1. Boundary state and boundary two point function,''
[arXiv:hep-th/0001012 [hep-th]].


\bibitem{Teschner:2000md}
J.~Teschner,
``Remarks on Liouville theory with boundary,''
PoS \textbf{tmr2000}, 041 (2000)
doi:10.22323/1.006.0041
[arXiv:hep-th/0009138 [hep-th]].

\bibitem{Lukyanov:2003nj}
S.~L.~Lukyanov, E.~S.~Vitchev and A.~B.~Zamolodchikov,
``Integrable model of boundary interaction: The Paperclip,''
Nucl. Phys. B \textbf{683}, 423-454 (2004)
doi:10.1016/j.nuclphysb.2004.02.010
[arXiv:hep-th/0312168 [hep-th]].

\bibitem{Sen:1998sm}
A.~Sen,
``Tachyon condensation on the brane anti-brane system,''
JHEP \textbf{08}, 012 (1998)
doi:10.1088/1126-6708/1998/08/012
[arXiv:hep-th/9805170 [hep-th]].

\bibitem{Smith:1991up}
E.~Smith and J.~Polchinski,
``Duality survives time dependence,''
Phys. Lett. B \textbf{263}, 59-62 (1991)
doi:10.1016/0370-2693(91)91707-3


\bibitem{Ginsparg:1986wr}
P.~H.~Ginsparg and C.~Vafa,
``Toroidal Compactification of Nonsupersymmetric Heterotic Strings,''
Nucl. Phys. B \textbf{289}, 414 (1987)
doi:10.1016/0550-3213(87)90387-7


\bibitem{Buscher:1987sk}
T.~H.~Buscher,
``A Symmetry of the String Background Field Equations,''
Phys. Lett. B \textbf{194}, 59-62 (1987)
doi:10.1016/0370-2693(87)90769-6



\bibitem{Itzhaki:2021scf}
N.~Itzhaki,
``String Theory and The Arrow of Time,''
JHEP \textbf{03}, 192 (2021)
doi:10.1007/JHEP03(2021)192
[arXiv:2101.10142 [hep-th]].


\end{thebibliography}
\end{document}